\documentclass[%
 aip,
 jcp,
 amsmath,amssymb,
reprint,%
]{revtex4-1}

\usepackage{graphicx}
\usepackage{dcolumn}
\usepackage{bm}
\usepackage{multirow}

\newcommand{\ket}[1]{\left|#1\right\rangle}

\begin{document}


\title{Laser-induced fluorescence studies of HfF$^+$ produced by autoionization}

\author{Huanqian Loh}
\email{loh@jilau1.colorado.edu}
 \affiliation{JILA, National Institute of Standards and Technology and University of Colorado, and Department of Physics, University of Colorado, Boulder, Colorado 80309-0440, USA}
\author{Jia Wang}
\affiliation{JILA, National Institute of Standards and Technology and University of Colorado, and Department of Physics, University of Colorado, Boulder, Colorado 80309-0440, USA}
\author{Matt Grau}
\affiliation{JILA, National Institute of Standards and Technology and University of Colorado, and Department of Physics, University of Colorado, Boulder, Colorado 80309-0440, USA}
\author{Tyler S. Yahn}
\affiliation{JILA, National Institute of Standards and Technology and University of Colorado, and Department of Physics, University of Colorado, Boulder, Colorado 80309-0440, USA}
\author{Robert W. Field}
\affiliation{Department of Chemistry, Massachusetts Institute of Technology, Cambridge, Massachusetts 02139, USA}
\author{Chris H. Greene}
\affiliation{JILA, National Institute of Standards and Technology and University of Colorado, and Department of Physics, University of Colorado, Boulder, Colorado 80309-0440, USA}
\author{Eric A. Cornell}
\email{ecornell@jilau1.colorado.edu}
\affiliation{JILA, National Institute of Standards and Technology and University of Colorado, and Department of Physics, University of Colorado, Boulder, Colorado 80309-0440, USA}

\date{\today}

\begin{abstract}
Autoionization of Rydberg states of HfF, prepared using the optical-optical double resonance (OODR) technique, holds promise to create HfF$^+$ in a particular Zeeman level of a rovibronic state for an electron electric dipole moment (eEDM) search. We characterize a vibronic band of Rydberg HfF at 54~cm$^{-1}$ above the lowest ionization threshold and directly probe the state of the ions formed from this vibronic band by performing laser-induced fluorescence (LIF) on the ions. The Rydberg HfF molecules show a propensity to decay into only a few ion rotational states of a given parity and are found to preserve their orientation qualitatively upon autoionization. We show empirically that we can create 30\% of the total ion yield in a particular $\ket{J^+, M^+}$ state and present a simplified model describing autoionization from a given Rydberg state that assumes no angular dynamics.
\end{abstract}

\maketitle


\section{Introduction}

Trapped HfF$^+$ has been identified as a promising candidate for testing fundamental symmetries and extensions to the Standard Model through an electron electric dipole moment (eEDM) search \cite{LBL10, MBD06, PMI07, PMT09}. A high-precision search for the eEDM demands the preparation of as many HfF$^+$ ions of a single isotope as possible in a particular $\ket{J^+, M^+}$ rovibronic, Zeeman sublevel. Unwanted ions created in other states are co-trapped with and can collide with the relevant ions, contributing to the dephasing of the latter species. The reduction in both the relevant sample number and coherence time can doubly hurt the sensitivity of an eEDM experiment. Hence, the state preparation of HfF$^+$ in a single rovibronic, Zeeman level is an important but non-trivial task.

The strategy we use to prepare HfF$^+$ in a state-selective manner is the autoionization of Rydberg HfF molecules excited from ground state HfF using the optical-optical double resonance (OODR) technique \cite{FLB04}. These autoionizing states lie above the lowest ionization threshold. Their ion-core is excited and decays after a short time, releasing energy to kick out the Rydberg electron. Will the resultant state of the ion have some memory of the state of its parent Rydberg molecule?

In principle, the above question can be answered by building a model of autoionization based on multichannel quantum defect theory (MQDT) \cite{Fano70, JA77, Seaton83, GJ85}, like that performed for CaF \cite{KCW11}. This model calculates the quantum defect matrix elements $\mu(R,\mathcal{E})$ and their derivatives with respect to the internuclear distance $R$ and energy $\mathcal{E}$, which can then be used to describe the Rydberg energy spectrum, dynamics of autoionization and resultant ion states. However, constructing such a quantum defect model demands extensive knowledge of Rydberg levels to provide input parameters; such knowledge is presently lacking for HfF.

Alternatively, we tackle the question of Rydberg state -- ion state branching ratios by experimentally probing the post-autoionization states of HfF$^+$ with laser-induced fluorescence (LIF). Ion LIF has been reported before by other groups as a tool to examine the states of ions created from resonance-enhanced multiphoton ionization (REMPI) \cite{DM87, FEI88, FEI89, XZ89, PDD81} , electron-impact ionization \cite{NNK87} and ion-molecule collisions \cite{GHL83}. A slightly different but related form of state detection is the grating-dispersed fluorescence obtained from highly-excited ions formed by photoionization using synchrotron light sources \cite{KHE87, PDP87, KCP92, DWM94, PCR95}. In our experiment, the ions are formed from a Rydberg band at 54~cm$^{-1}$ above the lowest ionization threshold. The upper levels of this Rydberg band are energetically allowed to decay only to the $X ^1\Sigma^+ (\nu^+ = 0)$ vibronic ground state of HfF$^+$, although the ions could be spread out over as many as $\approx 200$ distinct $\ket{J^+, M^+}$ rotational-Zeeman sublevels. The ion distribution over the various $\ket{J^+, M^+}$ states remain to be unveiled through their LIF intensities.

To predict the ion population distributions, which are connected to the observed fluorescence intensities, we present a simplified model of autoionization that assumes: 1.~the Rydberg molecule has the same electronic ion-core as its autoionization product, and 2.~the Rydberg electron flies off with the same angular momentum as it possessed when it was bound to the molecule, i.~e.~the dynamics of autoionization are radial only. The first assumption is based on the picture that the Rydberg state undergoes vibrational autoionization as opposed to electronic autoionization; in vibrational autoionization, propensity rules tend to favor a $\nu_{Ryd} = 1 \rightarrow \nu^+ = 0$ decay process \cite{FLB04}. The second assumption means that unlike the quantum defect model, the simplified model neglects the matrix elements of $\partial \mu/\partial R$ that are off-diagonal in the Rydberg electron's orbital angular momentum $l$. We do not yet understand the HfF Rydberg spectrum sufficiently well, in fact, to independently test these two assumptions. Instead, the severe lack of information obligates us to propose an initial model that is as simple as possible. As will be seen, the model can be adjusted to give a good account of the ion rotational distributions, and with no further adjustment it does well predicting the orientation and $M^+$ populations of a particular $J^+$ level. That said, on the basis of the present work alone, we cannot rule out electronic autoionization nor the presence of angular dynamics in the autoionization process. The remaining sections of this paper detail the experimental methods, autoionization theory and results of our LIF studies on autoionized HfF, with the primary goal toward maximizing ion creation in a desired single $\ket{J^+, M^+}$ state.


\section{Experiment}
\label{sec:expt}

\begin{figure}
\includegraphics{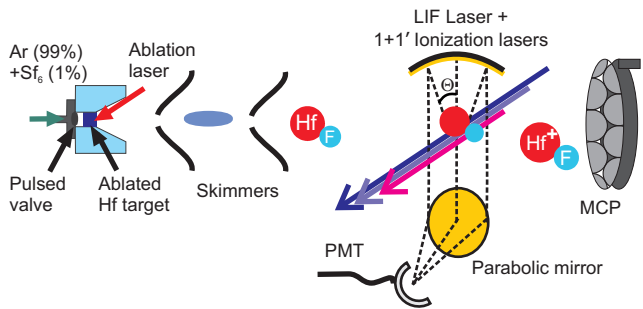}
\caption{(Color online) Schematic of the OODR-LIF apparatus. The molecular beam axis, the direction of propagation of the three co-propagating lasers and the axis of fluorescence photon collection are mutually perpendicular. From the ions, the fluorescence collection mirror on top subtends a polar angle of $\delta\theta = 75^\circ$, whereas the parabolic mirror below subtends $\theta = -11.8 ^\circ$ to $8.4^\circ$. A 820nm-bandpass filter, placed between the top collection mirror and the parabolic mirror, is not shown.}
\label{fig:ionlifsetup}
\end{figure}

Fig.~\ref{fig:ionlifsetup} shows a schematic of the OODR-LIF experiment setup used to ionize a HfF molecular beam and probe the resultant ionic states. The source and OODR-LIF chambers are separately pumped. A gas comprising 1\% SF$_6$ and 99\% Ar is released into the source chamber through the opening of a pulsed valve (800~$\mu$m orifice, 120~psi backing pressure) for 150~$\mu$s. In the presence of SF$_6$, the ablation of a Hf rod by the fundamental of a Nd:YAG pulsed laser (5--7~ns, 25~mJ/pulse, focused beam diameter 230~$\mu$m) creates HfF. We can also use the second harmonic of the Nd:YAG laser to perform the ablation; both cases gave no difference to the number of HfF molecules created. The HfF molecular beam is cooled through supersonic expansion to a rotational temperature of $\sim$ 10 K. The beam is collimated by two skimmers (3~mm orifice diameter, separation distance 12.5~cm) before it enters the OODR-LIF chamber. Three co-propagating lasers --- two for OODR and one for LIF of the ions --- enter and exit the OODR-LIF chamber through Brewster-angled windows mounted on 15-cm long tubes, in which baffles are placed to reduce scattered light. The three lasers intersect the molecular beam axis at 90$^\circ$ and the HfF$^+$ ions produced are accelerated by an electric field of 0.3~V/cm through 5~cm to two microchannel plates (MCP) stacked in a chevron configuration. The ion signal, after amplification by a transimpedance amplifier, is monitored on an oscilloscope. Fluorescence photons are collected above and below the ion beam for 2~$\mu$s after the ions are excited. The mirror on top collects most of the fluorescence photons and directs them to a parabolic mirror below, which in turn focuses them onto a photomultiplier tube (PMT). To further reduce scattered light counts, the photomultiplier tube is not gated on until 200~ns after the LIF laser fires. A bandpass filter centered at 820~nm is also placed between the top collection mirror and the parabolic mirror. The photon signals are amplified by a transimpedance amplifier, accumulated over at least 500 shots using a separate channel of the oscilloscope and counted individually using a peak-finding algorithm. 

\begin{figure}
\includegraphics{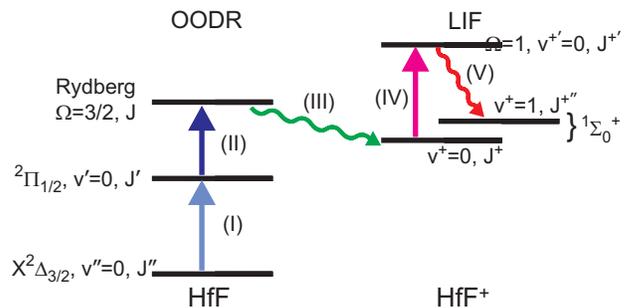}
\caption{(Color online) Neutral HfF and HfF$^+$ energy levels addressed in the OODR-LIF scheme. The quantum numbers assigned to each rotational level follow those given in Section~\ref{sec:theory}. The transitions marked by roman numbers correspond to the following: I) first excitation photon in the OODR technique, II) second excitation photon in the OODR technique, III) autoionization, IV) laser-driven transition in LIF, and V) fluorescence photons detected by the PMT.}
\label{fig:oodrlifscheme}
\end{figure}

All three lasers involved in addressing the energy levels in the OODR-LIF scheme, as depicted in Fig.~\ref{fig:oodrlifscheme}, are dye lasers. The first OODR laser pulse (1.5~$\mu$J/pulse, 10~ns, 150~MHz FWHM) is the output of a home-built two-stage Rhodamine 101 dye cell amplifier, frequency doubled with a $\beta$-BBO crystal. The dye cell amplifier is seeded by a continuous-wave ring dye laser operating with Rhodamine 610 Chloride dye and pumped by the second harmonic of a Nd:YAG laser. The Nd:YAG second harmonic at 532~nm also pumps a dye laser operating with Pyridine 2, the output of which is frequency-doubled with a KDP crystal to produce the second OODR laser pulse (100--200~$\mu$J/pulse, 10~ns, 0.1~cm$^{-1}$ FWHM). The second laser pulse is delayed by 14~ns relative to the first laser. The LIF laser pulse (6--8~ns, 0.5~uJ/pulse, 100--200~MHz) is the output of a home-built two-stage LDS 798 dye cell amplifier, which is pumped by the second harmonic of a Nd:YAG laser and seeded by an external cavity diode laser at 770~nm to address the [13.0]$1 (\nu' = 0) \leftarrow X^1\Sigma^+ (\nu'' = 0)$ transition \footnote[1]{``[13.0]$1(\nu' = 0)$'' refers to an excited state of vibrational quantum number $\nu' = 0$, total angular momentum projected on the body-fixed axis $\Omega = 1$, and at 13,0xx cm$^{-1}$ from the vibronic ground state of HfF$^+$.}$^, $ \cite{Cossel11} in $^{180}$Hf$^{19}$F$^+$. The LIF laser is delayed by 2~$\mu$s relative to the OODR lasers. Dichroic mirrors are used to overlap the three laser beams spatially before they enter the vacuum chamber. The laser pulses are linearly polarized either by the orientation of their frequency-doubling crystals or by passing through a polarizing beam splitter. They can then be set to either left or right circular polarizations using quarter and half waveplates for the appropriate wavelength ranges. Throughout the experiment, the frequencies of the seed lasers for both the first OODR and LIF lasers are locked to a single rotational line using a high-precision wavemeter that is regularly calibrated against the $^{87}$Rb $D_2$ transition. The frequency of the second OODR laser is monitored by both the high-precision wavemeter and a second wavemeter that is internally calibrated against Ne spectral lines.

To record OODR autoionization spectra, we fix the first OODR laser pulse on a single rotational line of a given parity of the [31.5]$1/2 \leftarrow X^2\Delta_{3/2}$ transition \footnote[2]{Similarly, ``[31.5]$1/2$'' refers to an excited state of $\Omega = 1/2$, and at 31,5xx cm$^{-1}$ from the vibronic ground state of HfF.}$^, $  \cite{LLS11} in the $^{180}$Hf$^{19}$F isotope while scanning the second laser pulse in frequency. We use circular dichroism techniques (i.e.~we compare the ion signal obtained when both lasers are polarized with the same helicity versus when polarized with opposite helicity) to identify transitions to various Rydberg rotational levels (denoted as $J$) in the autoionization spectra \cite{ZKZ04, PF08}.

Rovibronic state detection of the HfF$^+$ ion yield is accomplished by counting fluorescence photons that are emitted down to the $X^1\Sigma^+ (\nu'' = 1)$ vibronic level \cite{PMT09, BAB11} when the ions are excited by the LIF laser pulse. The frequency of the LIF laser is chopped every 100 shots between being on-resonance and 500~MHz off-resonance of a rotational line. The number of fluorescence photons detected when the horizontally-polarized LIF laser is tuned to a R(0), Q(1), Q(2), ... , Q(5) transition is related to the rotational populations in the $J^+ = 0, 1, 2, ..., 5$ levels respectively. Both OODR laser pulses are set to right circular polarizations. LIF detection is carried out when the second OODR laser pulse is tuned to both on-resonance and $\pm$0.35~cm$^{-1}$ off-resonance of an autoionizing line, so as to subtract out the contribution to the measured LIF intensities from non-resonantly produced ions. 

To determine the orientation of the ions formed in the $J^+ = 1$ rotational state, the LIF laser is fixed on either the Q(1) or the R(1) transition. With both OODR lasers right circularly polarized, we chop between having the LIF laser left and right circularly polarized and measure the respective fluorescence signals. Half waveplates and quarter waveplates are mounted on motorized rotation stages to perform the polarization switching.

\section{Autoionizing States of Hafnium Fluoride}

\begin{figure}
\includegraphics{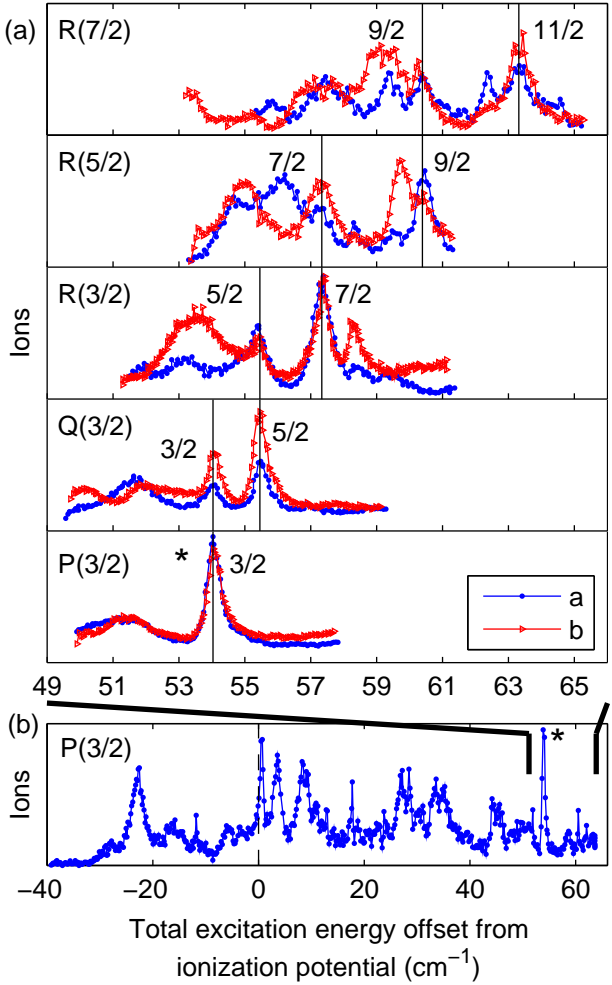}
\caption{(Color online) (a) Stacked plots of OODR autoionization spectra, measured by scanning the frequency of the second photon while holding the first photon fixed on different ground-intermediate transitions (denoted on the left of the figure as `P(3/2)', etc.) to access different intermediate rotational states. The identified Rydberg rotational energy levels are marked by solid black lines with their corresponding assignments. The autoionization spectra shown as blue dots (red triangles) belong to the same parity `$a$' (`$b$'). The ionization energy given by the $x$-axis is referenced to the ground rovibronic level in neutral HfF \cite{AHT04} and is offset from the ionization potential, 59462(2)~cm$^{-1}$. (b) A broader scan of OODR ionization spectra is conducted in an electric field of $\approx 25$ V/cm, which explains the appearance of peaks in ion creation at $\sim 30$ cm$^{-1}$ below the ionization potential of HfF. The peak marked by an asterisk is the same peak as that in (a).}
\label{fig:plotcfions}
\end{figure}

Transitions to different Rydberg rotational levels can be made by tuning the first OODR laser pulse to access intermediate states of different $J'$ and parity, as shown by the stacked plots in Fig.~\ref{fig:plotcfions}a. The various rotational transitions driven by the first OODR laser pulse are labeled on the top left corner of each subplot. Since the $\Omega$-doublets of the $X^2\Delta_{3/2}$ ground state in HfF cannot be resolved by the first OODR laser, the parity of the Rydberg states can only be determined up to an overall sign and is assigned as either `$a$' or `$b$' instead of as `e' or `f'. The Rydberg rotational levels are identified using the circular dichroism technique outlined in Section~\ref{sec:expt}; for the autoionization spectrum shown in Fig.~\ref{fig:plotcfions}a, both OODR lasers are chosen to be right circularly polarized to enhance the autoionization line intensity from higher $J$ states. The level spacings arise from half-integer pattern-forming quantum numbers, which indicate that the Rydberg electron is core-penetrating, i.e.~the inner lobe of its wavefunction lies inside the molecular ion-core \cite{KBC04}. Since no transition to the $J = 1/2$ level is observed, the Rydberg vibronic band is inferred to have $\Omega = 3/2$. The $\Omega$-doublets of the Rydberg vibronic band have no observable energy splitting, even up to $J = 11/2$. The Rydberg rotational energies are fit to the polynomial $E(J) = T_0 + B_{Ryd} J (J + 1)$, yielding a rotational constant of $B_{Ryd} = 0.2911(6)$ cm$^{-1}$. The $J = 7/2$ and $11/2$ levels are found to be perturbed by as much as -0.16(2) cm$^{-1}$ and +0.27(3) cm$^{-1}$ from the expected energy levels of a rigid rotor. 

The Rydberg vibronic band characterized at 54~cm$^{-1}$ above the lowest ionization threshold (59462(2) cm$^{-1}$) is part of a broader scan of the autoionization spectrum, shown in Fig.~\ref{fig:plotcfions}b. In contrast to the anomalously clean series of rotational lines in Fig.~\ref{fig:plotcfions}a, the HfF autoionization spectrum is generally cluttered. Although the pulse energy of the second OODR laser is reduced to avoid saturating the transitions to Rydberg states, the autoionization spectrum contains many broad features, the narrowest of which, at 0.3~cm$^{-1}$, is three times broader than the linewidth of the second OODR laser. This broadening cannot be attributed to either lifetime or hyperfine structure. The clutter of features makes it impossible for us to identify a clean Rydberg series of lines leading to ionization thresholds \footnote[3]{The lowest and excited HfF ionization thresholds have been identified using pulsed field ionization -- zero electron kinetic energy (PFI-ZEKE) photoelectron spectroscopy\cite{BAB11}}.

The identification of various Rydberg series and their underlying Rydberg electron character could potentially lead to predictions of ion rovibronic distributions. Alternatively, we can measure the distributions of the ions formed and work backwards to elucidate the character of the Rydberg electron. Toward this end, we developed a simplified model of autoionization, presented in the following section.

\section{Theory}
\label{sec:theory}

In this simplified model of autoionization, the Rydberg molecule is treated as a Rydberg electron attached to a $^1\Sigma^+$ ion-core at relatively short distances. The good quantum numbers of the resonant Rydberg state include the total angular momentum of the molecule $J$, its projection on the molecular $z$-axis $\Omega$ (approximately), its projection on the laboratory $z$-axis $M_J$, and the parity of the molecule. The quantization ($\hat{z}$) axis in the laboratory is provided by the direction of laser propagation. The anisotropic interaction between the outermost electron and the ion-core couples the orbital angular momentum $l$ and total angular momentum $j$ of the Rydberg electron. Neglecting the vibrational part, we write the short-range Rydberg states as a superposition of basis states having definite values of $l$ and $j$:
\begin{equation} \label{eqn:resryd}
\Psi _{res}  = \sum\limits_{lj} {A_{lj} \psi _{lj}^{\left| \Omega  \right|P}} \, ,
\end{equation}
where $A_{lj}$ is the probability amplitude of the corresponding $\{l, j\}$ partial wave, and where
\begin{eqnarray}\label{SRpartialwaves}
\psi _{lj}^{\left| \Omega  \right|P} = \frac{1}{{\sqrt 2 }}\left[ {\mathcal{R}\left( {J\left| \Omega  \right|M_J } \right)\sum\limits_{\Omega \sigma } {C_{l\Omega ;s\sigma }^{j\left| \Omega  \right|} Y_{l\Omega } \chi _{s\sigma } }}\right . \nonumber \\ 
\left.{ - \left(  -1  \right)^{j + J - l+P  } \mathcal{R}\left( {J - \left| \Omega  \right|M_J } \right)\sum\limits_{\Omega \sigma } {C_{l\Omega ;s\sigma }^{j - \left| \Omega  \right|} Y_{l\Omega } \chi _{s\sigma } } } \right] \, .
\end{eqnarray}
In Eq. (\ref{SRpartialwaves}), $C, Y_{l\Omega}$, and $\chi_{s\sigma}$ denote a Clebsch-Gordan coefficient, spherical harmonic, and spinor wavefunction, respectively. $\mathcal{R}\left( {J \Omega M_J } \right)$ is related to the Wigner rotation matrix $D_{M_J \Omega}^J$, which is a function of the Euler angles $\{\alpha, \beta, \gamma\}$ ,
\begin{equation}
\mathcal{R}\left( {J \Omega M_J } \right) = \sqrt {\frac{{2J + 1}}{{8\pi ^2 }}} \left[ {D_{M_J \Omega }^J \left( {\alpha \beta \gamma } \right)} \right]^* \, .
\end{equation}
The parity of the molecule is given by $\left ({-1}\right) ^ {P}$.

After autoionization, the distance between the emitted electron and the ion-core is large. This long-range (electron $+ \rm{HfF}^+$) system is described by the total ionic angular momentum $J^+$, its projection $M^+$ on the laboratory z-axis, and its projection $\Omega^+$ on the body-fixed z-axis. Here, $\Omega^+ = 0$ since the ion is in a $^1\Sigma^+$ state. Because the short-range states have a definite total angular momentum $J$ and projection on the laboratory z-axis $M_J$, we also want to construct the long-range states with definite $J$ and $M_J$. We assume that the Rydberg electron flies off with the same $\{l, j\}$ angular momenta it possessed when bound to the molecule, i.e.~there are no angular dynamics in autoionization. The $\{l, j\}$ partial waves of the system after autoionization can then be written as
\begin{eqnarray}\label{LRpartialwaves}
\phi _{lj}^{J^+ } &=& \sum\limits_{M^+  m_j } {C_{J^+  M^+  ;jm_j }^{JM_J } \mathcal{R}\left( {J^+ , \Omega^+ = 0, M^+ } \right)} \nonumber \\ 
&& \times \sum\limits_{\lambda m_s } {C_{l\lambda ;sm_s }^{jm_j } Y_{l\lambda } \chi _{sm_s } } \, .
\end{eqnarray}
Eqs.~(\ref{SRpartialwaves}) and (\ref{LRpartialwaves}) can be related by a rotational frame transformation,
\begin{eqnarray}
\psi _{lj}^{\left| \Omega  \right|P}  = {\left(  -1  \right)^{j + \left| \Omega  \right|} }\sum\limits_{J^+} {C_{j - \left| \Omega  \right|,J\left| \Omega  \right|}^{J^+  0} } \sqrt {1 + \left(  -1  \right)^{l + J^+   + P} } \phi _{lj}^{ J^+} \, . \nonumber \\
\end{eqnarray}
Hence, after autoionization, the probability for a Rydberg state with quantum numbers $J$ and $M_J$ to produce an ion in the $\ket{J^+, M^+}$
 state is
\begin{eqnarray} \label{eqn:autoionizationcg}
P_{J^+,M^+  }^{J,M_J }  = &\sum\limits_{lj}& \left| {A_{lj} } \right|^2 \left[ {C_{j - \left| \Omega  \right|, J \left| \Omega  \right|}^{J^+  0} } \right]^2 \left[ {C_{J^+  M^+,  jm_j }^{JM_J } } \right]^2 \nonumber \\
&& \times \left[ {1 + \left(  -1  \right)^{l + J^+   + P} } \right] \, .
\end{eqnarray}

In the experiment determining the ions' rotational distribution, the resonant Rydberg state is prepared by two right circularly polarized laser pulses. The OODR excitation steps are denoted as $(X ^2\Delta_{3/2}) J'' \rightarrow (\Pi_{1/2}) J' \rightarrow$ (Rydberg $\Omega = 3/2$) $J$. The molecules are assumed to reside initially in a random distribution of $M_J''$ sub-levels. The relative probability for a resonant Rydberg molecule to be prepared in a $\ket{J, M_J}$ level is then given by
\begin{equation} \label{eqn:rydstateprep}
P_{M_J }^J  = \left| {C_{J''M_J  - 2;11}^{J'M_J  - 1} C_{J'M_J  - 1;11}^{JM_J } } \right|^2 \, ,
\end{equation}
where saturation effects are assumed negligible. The final (unnormalized) rotational distribution of the ions is obtained by combining Eq.~(\ref{eqn:autoionizationcg}) with Eq.~(\ref{eqn:rydstateprep}):
\begin{equation} \label{eqn:ionrotpop}
P_{J^+  M^+  }^J  = \sum\limits_{M_J } {P_{M_J }^J P_{J^+  M^+  }^{M_J } } \, .
\end{equation}

The rotational distribution given by Eq.~(\ref{eqn:ionrotpop}), however, cannot be compared against the experiment directly. In the experiment, LIF signals for the different $J^+$ states are measured. The ions are excited by an x-polarized laser tuned to the rotational transition $J^+ \rightarrow J^{+'}$ and subsequently emit photons as they decay radiatively to some lower state $J^{+''}$. The x-polarized laser can be described by a linear combination of two spherical harmonics $\frac{1}{{\sqrt 2 }}\left( {Y_{1 -1}  - Y_{11} } \right)$. Since the photomultiplier tube collects photons of both $Y_{1 -1}$ and $Y_{11}$ polarizations emitted in almost all directions, we can add the probabilities corresponding to each spherical harmonic incoherently. For a given spherical harmonic $Y_{1q}$, the relative signal strength has been adapted as a generalization of Eq. (23) in Ref.~\citenum{HW05}:
\begin{eqnarray}
&& I_{J^+ M^+, J^{+'}}^ {(q)} \nonumber \\
&=& \left( {2J^+   + 1} \right)\left( {2J^ {+'}   + 1} \right) \nonumber \\ 
&\times& \left( {\begin{array}{*{20}c}
   {J^+  } & 1 & {J^ {+'}  }  \\
   { - \Omega^+  } & {\Omega^{+'}   - \Omega^+  } & {\Omega^{+'}  }  \\
\end{array}} \right)^2 \left( {\begin{array}{*{20}c}
   {J^+  } & 1 & {J^{+'}  }  \\
   { - M^+  } & { - q} & {M^{+'}  }  \\
\end{array}} \right)^2 \nonumber \\
&\times& \sum\limits_{J^{+  ''} = \left| {J^{+'} - 1} \right|}^{J^{+'} + 1} {\sum\limits_{M^{+  ''} =  - J^{+  ''}}^{J^{+  ''}} {\left( {2J^{+  '} + 1} \right)\left( {2J^{+  ''} + 1} \right)} } \nonumber \\
&\times& \left( {\begin{array}{*{20}c}
   {J^{+'}} & 1 & {J^{+  ''}}  \\
   { - \Omega^{+'}} & {\Omega^{+'} - \Omega^{+  ''}} & {\Omega^{+  ''}}  \\
\end{array}} \right)^2\left( {\begin{array}{*{20}c}
   {J^{+'}} & 1 & {J^{+  ''}}  \\
   { - M^{+'}} & { - q'} & {M^{+  ''}}  \\
\end{array}} \right)^2 \, . \nonumber \\
\end{eqnarray}
The first two lines describe the laser excitation step and the last two lines describe the spontaneous emission step of the LIF process. The summation in the last two lines gives unity. Hence, for a given $\ket{J^+, M^+}$ state, the relative LIF signal strength for an x-polarized LIF excitation is given by
\begin{eqnarray}
I_{J^+  M^+, J^{+'}} && = \frac{1}{2}\left({I_{J^+  M^+, J^{+'}  }^{(+1)}+I_{J^+  M^+, J^{+'}  }^{(-1)}}\right) \nonumber \\
&& = \frac{1}{2} \left( {2J^+   + 1} \right)\left( {2J^{+'}   + 1} \right) \nonumber \\
\times && \left( {\begin{array}{*{20}c}
   {J^+  } & 1 & {J^{+'}}  \\
   { - \Omega^+  } & {\Omega^+   - \Omega^{+'}} & {\Omega^{+'}}  \\
\end{array}} \right)^2 \nonumber \\ 
\times && \left[ {\left( {\begin{array}{*{20}c}
   {J^+  } & 1 & {J^{+'}}  \\
   { - M^+  } & { - 1} & {M^{+'}}  \\
\end{array}} \right)^2  + \left( {\begin{array}{*{20}c}
   {J^+  } & 1 & {J^ {+'}}  \\
   { - M^+  } & 1 & {M^ {+'}}  \\
\end{array}} \right)^2 } \right]. \nonumber \\
\end{eqnarray}
Finally, the relative LIF signal strength, to be compared against the measured LIF intensities that trace over the $M^+$ levels, is given by
\begin{equation}\label{eqn:LIFsignal}
I_{J^+, J^{+'}  }  = \frac{1}{\mathcal{N}}\sum\limits_{M^+  } {P_{J^+  M^+  }^J I_{J^+  M^+, J^{+'}  } } \, ,
\end{equation}
where $\mathcal{N}$ is a normalization factor chosen such that $\sum\limits_{\{J^+, J^{+'}\}}I_{J^+, J^{+'}} = 1$. We use Eq. (\ref{eqn:LIFsignal}) to fit $I_{J^+, J^{+'}}$ to the measured LIF signal and determine the values of $\left| {A_{lj} } \right|^2$. The effects of excited state $(J^{+'})$ alignment and orientation on the photon collection efficiency were evaluated and found to be very small.

When probing the orientation of the ions, the LIF excitation laser is either right or left circularly polarized. The LIF signals for these cases are then given by
\begin{subequations} \label{eqn:LIFLRsignal}
\begin{equation}
\gamma_R = \sum\limits_{M^+ } P_{J^+  M^+  }^{J} I_{J^+  M^+  ,J^{ + '} }^{\left( 1 \right)} \, ,
\end{equation}
\begin{equation}
\gamma_L =  \sum\limits_{M^+ } P_{J^+  M^+  }^{J} I_{J^+  M^+  ,J^{ + '} }^{\left( { - 1} \right)} \, .
\end{equation}
\end{subequations}


\section{Rotational Distributions and Parity Propensities in Autoionization Decay}

\begin{figure}
\includegraphics{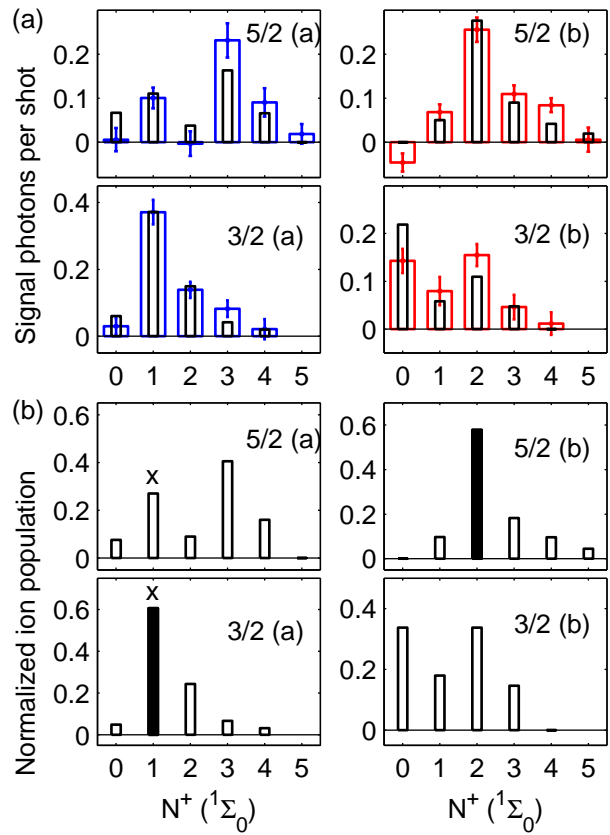}
\caption{(Color online) (a) The Rydberg molecules, labeled by $J$ and parity on the top right corner, are observed to autoionize into only a few ion rotational levels. These rotational distributions are probed by measuring the number of LIF photons, as shown in blue/red with error bars. In one case ($J^+ = 0 \leftarrow J = 5/2$ ($b$)), the measured LIF intensity appears negative, which is a statistical artifact of the multiple signal subtractions performed to account for scattered light photons and the rotational distribution of background ions. The narrower bar plots are theory fits to the data. (b) The calculated rotational distribution of ions is obtained based on the fits to the rotational line intensities above. The creation of $\sim 60$\% of the ions in a single rotational level, as highlighted by the solid black bars, can be a significant advantage for future experiments. The ions formed in the $J^+ = 1$ state, marked by crosses above the bar plots, have their $M^+$ distributions further examined (see Fig.~\ref{fig:orientation}).}
\label{fig:plotionlifdistrib}
\end{figure}

The fluorescence intensities, measured by exciting ions formed in various rotational states of the $X^1\Sigma^+, \nu^+ = 0$ vibronic ground state after autoionization, are given in Fig.~\ref{fig:plotionlifdistrib}a. We focus on the ions produced from both parities of the $J = 3/2$ and $5/2$ rotational Rydberg levels. Although the Rydberg molecules at 54~cm$^{-1}$ above ionization threshold are energetically allowed to decay into many rotational levels of HfF$^+$, they are observed to autoionize into only a few rotational levels. The Rydberg molecules of a given parity are further observed to decay into ion rotational levels of the same parity, i.e.~molecules from the `$a$' (`$b$') Rydberg state primarily form ions in the odd (even) rotational states, as shown on the left (right) side of Fig.~\ref{fig:plotionlifdistrib}a. Both the rotational propensity and parity propensity observations have also been reported for near-homonuclear molecular ions like N$_2^+$ and NO$^+$ created from REMPI \cite{FEI88, XZ89, PLZ96}, and are accounted for in a theory paper by Xie and Zare \cite{XZ90}. For a polar molecule like CaF$^+$, however, the parity propensity rule tends not to hold due to extensive $l$-mixing for the Rydberg electron. For the HfF vibronic band reported here, the parity propensity observation indicates the release of a Rydberg electron with $l$ of predominantly one parity. On the other hand, the non-negligible formation of ions in rotational states of both parities implies that the Rydberg electron was ejected with a superposition of orbital angular momenta. 

We use Eq.~(\ref{eqn:LIFsignal}) from the simplified autoionization model to generate fluorescence intensities for each allowed $\{l, j\}$ and fit them against the LIF data to obtain the angular momentum composition of the Rydberg electron. We note that since the overall parity of the Rydberg states cannot be determined \textit{a priori}, there are two possible sets of $A_{lj}$ to which the data could have fit. However, the fits only converged for one of the two cases, strongly suggesting that the Rydberg states denoted by `$a$' (`$b$') should be assigned the parity $P = +1 (-1)$ in Eq.~(\ref{eqn:resryd}). In this case, the fit results gave 67.5(3.7)\% $p_{3/2}$, 9.8(4.5)\% $d_{3/2}$ and 22.7(5.3)\% $d_{5/2}$ character to the Rydberg electron, where the numbers in parenthesis indicate the $1\sigma$-error. In accordance with the half-integer quantum number progression of the Rydberg rotational states, the Rydberg electron is found to be in primarily a core-penetrating state. The mixed $\{l, j\}$ character of the HfF Rydberg state has also been reported for CaF Rydberg molecules \cite{KCW11}. We acknowledge that in contrast to the presented model, other groups have reported significant angular dynamics in autoionization, which may be accounted for within the framework of MQDT: for example, the Rydberg electron may have only a single set of $\{l, j\}$ when bound to the molecule, but flies off with multiple $\{l, j\}$ \cite{PLZ96}, leading to the formation of ions in rotational states of both parities. Since the $\{l, j\}$ composition of the HfF Rydberg state is not known \textit{a priori}, we cannot claim the validity of our model over others.

From the fits to the LIF intensities, the population distribution of HfF$^+$ ions in the various rotational states can be obtained (see Fig.~\ref{fig:plotionlifdistrib}b). There can be as many as 60\% of the ions created in a single rovibronic state (highlighted as black solid bar plots), which is significant for the state selective creation of HfF$^+$ for future experiments.

\section{Preservation of Orientation in Autoionization Decay}

Certain experiments demand not only the creation of molecular ions in a particular rovibronic state, but in a single Zeeman level of that state. To this end, we experimentally determine the orientation of HfF$^+$ formed in the $J^+ = 1, \nu^+ = 0, X^1\Sigma^+$ state from the Rydberg levels $J = 3/2 (a)$ and $J = 5/2 (a)$ prepared using two right circularly polarized OODR photons. The `$a$' parity states are chosen to maximize the ion signal, given the parity propensity rule observed in the previous section. Unlike when measuring the ion rotational populations, the orientation was only determined for when the OODR lasers are tuned to the resonance of an autoionizing line, as the non-resonant ions were found to contribute to only $\approx 20$\% of the population in $J^+ = 1$.The orientation is related to the contrast ratio $\mathcal{C}$ for the fluorescence signal ($\gamma_R, \gamma_L$) measured when chopping between right and left circular polarizations for the LIF laser, where $\mathcal{C}$ is defined as
\begin{equation}
\label{eqn:contrastratio}
\mathcal{C} \equiv \frac{\gamma_R-\gamma_L}{\gamma_R+\gamma_L} \, .
\end{equation}
The contrast ratio is a convenient quantity immune to drifts in ion production. As a systematic check of our polarizations, we measured the contrast ratio when both OODR laser pulses are left circularly polarized and found $\mathcal{C}$ to be of the same magnitude but opposite sign as when both lasers were right circularly polarized, as expected.

\begin{figure}
\includegraphics{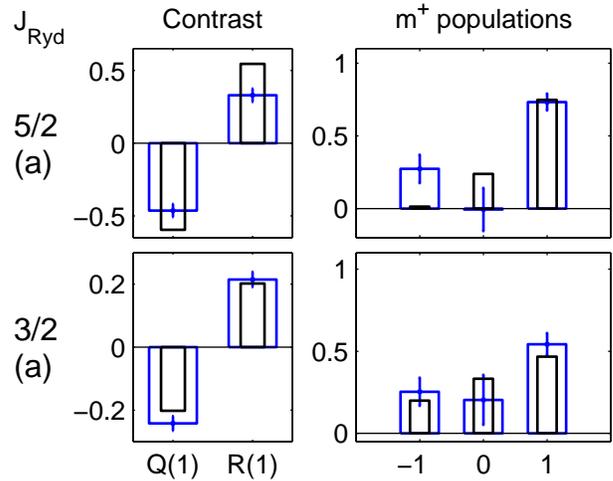}
\caption{(Color online) The Rydberg molecules in (top) $J = 5/2 (a)$ and (bottom) $J = 3/2 (a)$, prepared using two right circularly polarized photons in OODR, are observed to decay into the ion rotational level $J^+ = 1$ with a certain orientation. The ions' orientation, related to the $M^+$ population distribution (right), is inferred from measurements of contrast ratios (left) for the Q(1) and R(1) LIF transition. The narrower bar plots are the theory predictions for the $M^+$ populations.}
\label{fig:orientation}
\end{figure}

The contrast ratios for the Q(1) and R(1) transitions are related to the orientation $\mathcal{O}_0$ and alignment $\mathcal{A}_0$ of the ions through the following \cite{FM73}:
\begin{subequations} \label{eqn:CtoO0andA0}
\begin{equation}
\mathcal{C}^{Q(1)} = \frac{3 (1 + G_1) \mathcal{O}_0}{\mathcal{A}_0 - 2 (1 + G_2)}  \, ,
\end{equation}
\begin{equation}
\mathcal{C}^{R(1)} = \frac{15 (1 + G_3)\mathcal{O}_0}{\mathcal{A}_0 + 10 (1 + G_4)}  \, ,
\end{equation}
\end{subequations}
where $\{G_i\} \lesssim 0.05$ are correction factors that account for the anisotropy of the fluorescence collection setup (depicted in Fig.~\ref{fig:ionlifsetup}). The orientation and alignment parameters are in turn related to the $M^+$ populations, which are displayed as plots with error bars in Fig.~\ref{fig:orientation}. The narrower bar plots show the contrast ratios and $M^+$-level distributions predicted from Eq.~(\ref{eqn:LIFLRsignal}) of Section~\ref{sec:theory}, using the same Rydberg electron composition of $\{l, j\}$ inferred from the rotational distributions. The agreement between theory and experiment is good for the $M^+ = 1$ population but only fair for the $M^+ = 0, -1$ populations, part of which may be attributed to the oversimplification of the model's assumptions. We note that of the decay into a single rovibronic state, the population in a single Zeeman level may be as high as 54(7)\% (from $J = 3/2 (a)$) or even 73(6)\% (from $J = 5/2 (a)$). In particular, the former number, combined with the formation of 60\% of the ions in that same rovibronic state, means that as many as 30\% of the HfF$^+$ ions created from a certain autoionizing resonance can be in a single $\ket{J^+, M^+}$ level.

Different rotational levels can have different maximal (minimal) values for their orientation, given by $\mathcal{O}_0^{\mbox{\footnotesize{max}}} ( = -\mathcal{O}_0^{\mbox{\footnotesize{min}}}) = 1/(J+1)$. Therefore, instead of comparing the orientation parameter of the Rydberg molecule to that of the autoionized product, we compare the fractional orientation $\mathcal{O}_0'$:
\begin{equation}
\label{eqn:fracorient}
\mathcal{O}_0' = \frac{\mathcal{O}_0 - \mathcal{O}_0^{\mbox{\footnotesize{min}}}}{\mathcal{O}_0^{\mbox{\footnotesize{max}}} - \mathcal{O}_0^{\mbox{\footnotesize{min}}}} \, .
\end{equation}
$\mathcal{O}_0'$ is 1 (0) when only the $\ket{M^+ = +J^+}$ ($\ket{M^+ = -J^+}$) states are populated and 0.5 when there is no orientation. From the values reported in Table~\ref{tab:orientation}, we find that the autoionization of HfF Rydberg molecules to the $J^+ = 1, \nu^+ = 0, X^1\Sigma^+$ rovibronic state preserves orientation qualitatively, which bodes well for the creation of HfF$^+$ predominantly in a single Zeeman, rovibronic level for other experiments.

\begin{table}
\begin{center}
\caption{Fractional orientation $\mathcal{O}_0'$ of Rydberg HfF molecules in a given rotational level $J$ and of the ions formed in $J^+ = 1$ after autoionization. The HfF orientation is calculated from the polarizations of the OODR lasers; the HfF$^+$ orientation is predicted from the simplified model of autoionization described in the text; the HfF$^+$ (measured) column refers to values inferred from LIF polarization contrast ratios in Fig.~\ref{fig:orientation}.\label{tab:orientation}}
\begin{tabular}{|c|c|c|c|}
\hline \multirow{3}{*}{$J$} & \multicolumn{3}{|c|}{Fractional orientation}\\
\cline{2-4}  & Rydberg~HfF & HfF$^+$ & HfF$^+$ \\
& (calculated) & (predicted) & (measured) \\
\hline 3/2 ($a$) & 0.835 & 0.633 & 0.645(62) \\
 5/2 ($a$) & 0.865 & 0.867 & 0.730(86) \\
\hline
\end{tabular}
\end{center}
\end{table}


\section{Conclusions}

We have spectroscopically characterized a vibronic band of the autoionization spectrum in HfF at 54~cm$^{-1}$ above the lowest ionization threshold. We directly probe the decay of its Rydberg states ($J = 3/2, 5/2$) into various rotational states of the $X ^1\Sigma^+, \nu^+ = 0$ vibronic ground state of HfF$^+$ by performing laser-induced fluorescence on the ions. The measured fluorescence intensities are fit using a simplified model of autoionization that assumes no angular dynamics and that the Rydberg state has the same electronic ion-core as its autoionized state. The fit parameters imply that the Rydberg electron has dominant $p_{3/2}$ character with some mixing from the $d$ orbitals. Using the same Rydberg electron character, the model predicts qualitative preservation of orientation when the Rydberg molecule autoionizes to the $J^+ = 1, \nu^+ = 0, X^1\Sigma^+$ state, which was corroborated by fluorescence intensity measurements carried out after excitation by a circularly polarized laser. Thanks to a combination of rotational propensity, parity propensity and preservation of orientation during autoionization, we find that we can create as many as 30\% of the HfF$^+$ ions in a single Zeeman level of a rovibronic state. Having an initial population of 30\% of the ions in a single $\ket{J^+, M^+}$ level could prove to be very advantageous for future experiments such as the eEDM search.

We note that although this work has concentrated on one vibronic band of a core-penetrating Rydberg state, there is a body of core-nonpenetrating states in HfF that has yet to be uncovered with experiment. Core-nonpenetrating states have almost pure $l$ character. With $l \geq 2$ for HfF, the Rydberg electron hardly exerts a torque on the ion core rotational states when ejected, which means that the resultant ion rotational states are likely to follow that of the Rydberg molecules more closely. \cite{KayTBP} This lends core-nonpenetrating Rydberg states even more promise for populating ions of a desired $\ket{J^+, M^+}$ level with high efficiency.


%

\begin{acknowledgments} 
We thank R.~P.~Stutz for early contributions and helpful discussions. This work is funded by the NSF and the Marsico Research Chair. H.~Loh acknowledges support from A*STAR (Singapore).
\end{acknowledgments}


\end{document}